\documentclass[journal]{IEEEtran}

\usepackage{hyperref}
\pdfoutput=1
\ifCLASSINFOpdf
   \usepackage[pdftex]{graphicx}
\else
\fi
\begin{document}
%
\title{A Review of Wireless Body Area Networks \\ for Medical Applications}
%
%
%

\author{Sana~Ullah$^{\Psi}$, Pervez~Khan, Niamat~Ullah, Shahnaz Saleem, Henry~Higgins, and~Kyung Sup~Kwak
\IEEEcompsocitemizethanks{\IEEEcompsocthanksitem S.Ullah, P. Khan, N. Ullah, and K.S. Kwak are with the Graduate School of Telecommunication Engineering, Inha University, Incheon (402-751)
South Korea.
\IEEEcompsocthanksitem $\Psi$. Author to whom correspondence should be addressed E-mail: sanajcs@hotmail.com 
\IEEEcompsocthanksitem H. Higgins is with the Zarlink Semiconductor Company, Portskewett, Caldicot, United Kingdom 
\IEEEcompsocthanksitem S. Saleem is with the Graduate School of Computer Science and Engineering, Inha University, Incheon (402-751) South Korea. }
\thanks{Manuscript received March 8, 2009; revised May 16, 2009; published (International J. of Communications, Network and System Sciences (IJCNS), vol. 2, no. 8: 797-803.) July 27, 2009.}}
\maketitle

\begin{abstract}
Recent advances in Micro-Electro-Mechanical Systems (MEMS) technology, integrated circuits, and wireless communication have allowed the realization of Wireless Body Area Networks (WBANs). WBANs promise unobtrusive ambulatory health monitoring for a long period of time and provide real-time updates of the patient's status to the physician. They are widely used for ubiquitous healthcare, entertainment, and military applications. This paper reviews the key aspects of WBANs for numerous applications. We present a WBAN infrastructure that provides solutions to on-demand, emergency, and normal traffic. We further discuss in-body antenna design and low-power MAC protocol for WBAN. In addition, we briefly outline some of the WBAN applications with examples. Our discussion realizes a need for new power-efficient solutions towards in-body and on-body sensor networks.
\end{abstract}


%
\IEEEpeerreviewmaketitle

\section{Introduction}
\IEEEPARstart{C}{ardiovascular} disease is the foremost cause of death in the United States (US) and Europe since 1900. More than ten million people are affected in Europe, one million in the US, and twenty two million people in the world \cite{1}- \cite{3}. The number is projected to be triple by 2020. The ratio is 17\% in South Korea and 39\% in UK \cite{4}-\cite{5}. The healthcare expenditure in the US is expected to increase from \$2.9 trillion in 2009 to \$4 trillion in 2015 [6]. The impending health crisis attracts researchers, industrialists, and economists towards optimal and quick health solutions. The non-intrusive and ambulatory health monitoring of patient's vital signs with real time updates of medical records via internet provides economical solutions to the health care systems.

A WBAN contains a number of portable, miniaturised, and autonomous sensor nodes that monitor the body function for sporting, health, entertainment, and emergency applications. It provides long term health monitoring of patients under natural physiological states without constraining their normal activities. In-body sensor networks, especially, allow communication between implanted devices and remote monitoring equipments. They are capable of collecting information from Implantable Cardioverter Defibrillators (ICDs) in order to detect and treat ventricular tachyarrhythmia\footnote{Ventricular tachyarrhythmia are abnormal patterns of electrical activity originating within ventricular tissue} and to prevent Sudden Cardiac Death (SCD) \cite{7}. 

\begin{figure}[!h]
\centering
\includegraphics[width=2.5in]{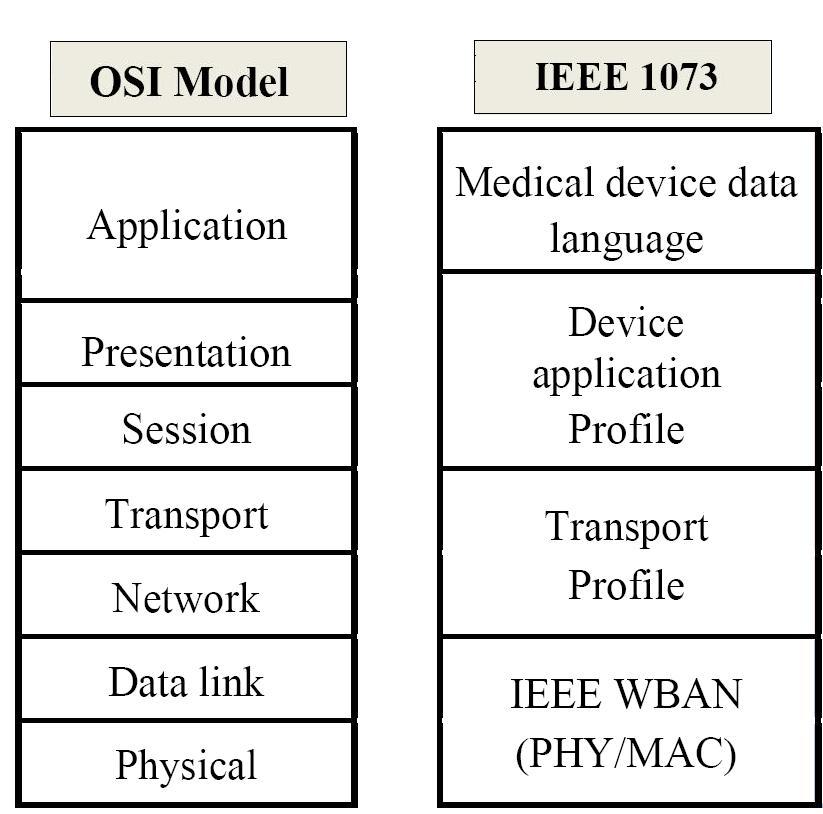}
\caption{Open Systems Interconnection (ISO) model and IEEE 1073}
\label{fig:1}
\end{figure}

A number of ongoing projects such as CodeBlue, MobiHealth, and iSIM have contributed to establish a proactive WBAN system \cite{8}-\cite{10}. A system architecture presented in \cite{11} performs real-time analysis of sensor's data, provides real-time feedback to the user, and forwards the user's information to a telemedicine server. UbiMon aims to develop a smart and affordable health care system \cite{12}. MIT Media Lab is developing MIThril that gives a complete insight of human-machine interface \cite{13}. HIT lab focuses on quality interfaces and innovative wearable computers \cite{14}. NASA is developing a wearable physiological monitoring system for astronauts called LifeGuard system \cite{15}. IEEE 802.15.6 aims to provide low-power in-body and on-body wireless communication standards for medical and non-medical applications \cite{16}. IEEE 1073 is working towards a seven layers solution for wireless communication in WBAN \cite{17}. Fig. \ref{fig:1} shows the IEEE 1073 model.

\begin{figure*}[!t]
\centering
\includegraphics[width=5in]{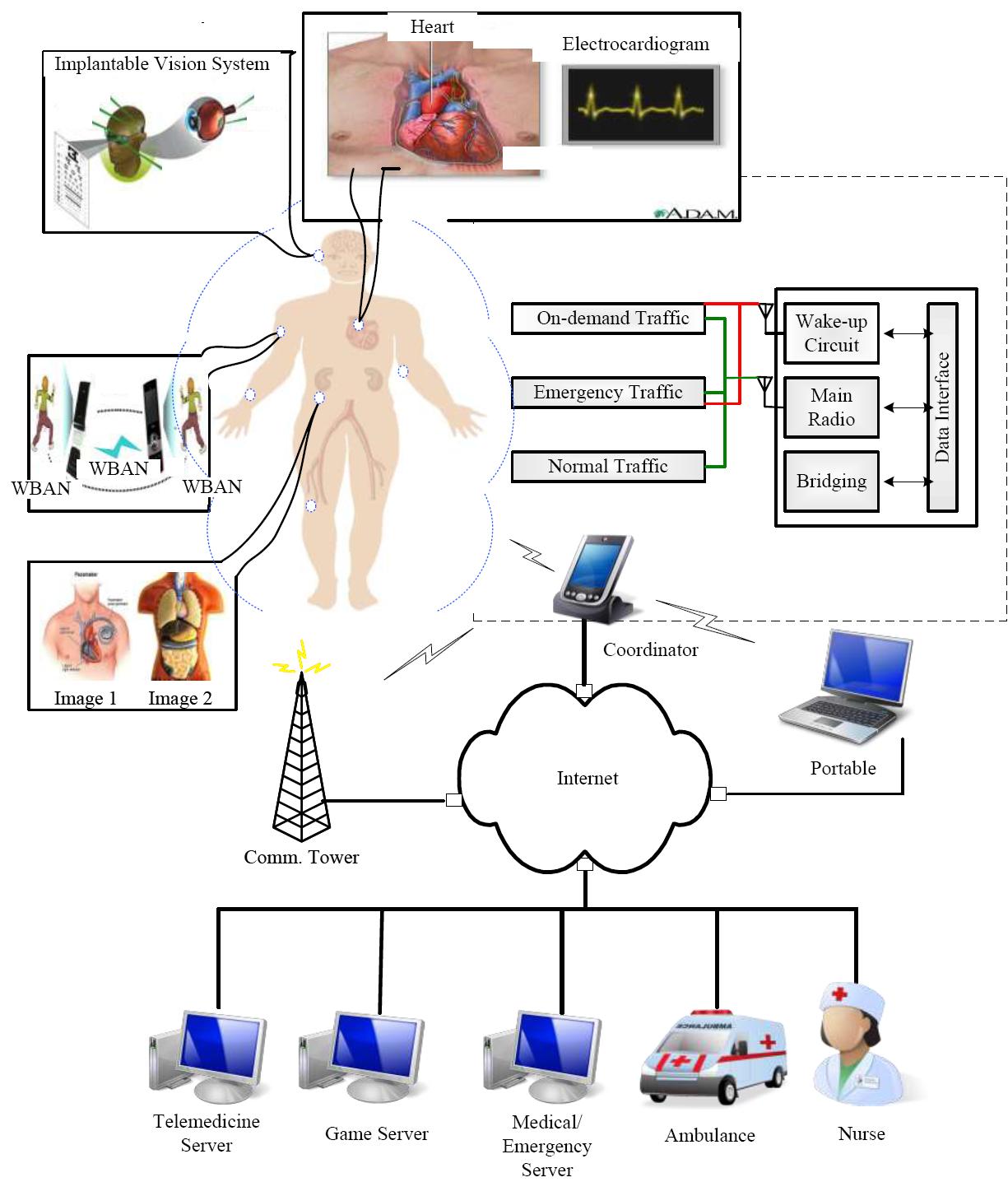}
\caption{A WBAN infrastructure for medical and non-medical applications}
\label{fig:2}
\end{figure*}

The rest of the paper is organized into five sections. Section 2 presents a WBAN infrastructure for medical and non-medical applications. Section 3 and Section 4 discuss in-body antenna design and low-power MAC protocol for WBAN. Section 5 outlines some of the WBAN applications. The final section concludes our work.

\section{WBAN Infrastructure}
A WBAN consists of in-body and on-body nodes that continuously monitor patient's vital information for diagnosis and prescription. Some on-body nodes can also be used for multimedia and gaming applications. A WBAN uses Wireless Medical Telemetry Services (WMTS), unlicensed Industrial, Scientific, and Medical (ISM), Ultra-wideband (UWB), and Medical Implant Communications Service (MICS) bands for data transmission. WMTS is a licensed band used for medical telemetry system. Federal Communication Commission (FCC) urges the use of WMTS for medical applications due to fewer interfering sources. However, only authorized users such as physicians and trained technicians are eligible to use this band. Furthermore, the restricted WMTS (14 MHz) bandwidth cannot support video and voice transmissions. The alternative spectrum for medical applications is to use 2.4 GHz ISM band that includes guard bands to protect adjacent channel interference. A licensed MICS band (402-405 MHz) is dedicated to the implant communication. 

\begin{table*}
\begin{center}
\caption{Body Electrical Properties \cite{19}}
\label{tab:1}
\begin{tabular}{|r|r|r|r|r|r|r|r|}
\hline
\multicolumn{1}{|r|}{Frequency (MHz)} &        \multicolumn{3}{|r|}{Muscle} &           \multicolumn{3}{|r|}{Fat} \\
\hline
\multicolumn{1}{|r|}{} & $\varepsilon_{r}$ & $\rho$ $(S.m^{-1})$ &   $Z_{0}(\Omega)$ & $\varepsilon_{r}$ & $\rho$ $(S.m^{-1})$ &  $Z_{0}(\Omega)$           \\
\hline
       100 &       66.2 &       0.73 &       31.6 &       12.7 &       0.07 &       92.4             \\
\hline
       400 &         58 &       0.82 &       43.7 &       11.6 &       0.08 &        108             \\
\hline
       900 &         56 &       0.97 &       48.2 &       11.3 &       0.11 &        111             \\
\hline
\end{tabular}
\end{center}
\end{table*}

Fig. \ref{fig:2} shows a WBAN infrastructure for medical and non-medical applications. As can be seen in the figure, the WBAN traffic is categorized into On-demand, Emergency, and Normal traffic. On-demand traffic is initiated by the coordinator or doctor to acquire certain information, mostly for the purpose of diagnostic recommendations. This is further divided into continuous (in case of surgical events) and discontinuous (when occasional information is required). Emergency traffic is initiated by the nodes when they exceed a predefined threshold and should be accommodated in less than one second. This kind of traffic is not generated on regular intervals and is totally unpredictable. Normal traffic is the data traffic in a normal condition with no time critical and on-demand events. This includes unobtrusive and routine health monitoring of a patient and treatment of many diseases such as gastrointestinal tract, neurological disorders, cancer detection, handicap rehabilitation, and the most threatening heart disease. The normal data is collected and processed by the coordinator. The coordinator contains a wakeup circuit, a main radio, and a bridging function, all of them connected to a data interface. The wakeup circuit is used to accommodate on-demand and emergency traffic. The bridging function is used to establish a logical connection between different nodes working on different frequency bands. The coordinator is further connected to telemedicine, game, and medical servers for relevant recommendations.

\section{In-body Antenna Design}
The band designated for in-body communication is MICS band, which is around 403MHz. The wavelength of this frequency in space is 744mm so a half wave dipole will be 372mm. Clearly, it is not possible to include an antenna of such dimensions in a body \cite{19}. These constraints make the available size much smaller than the optimum. Generally, the electrical properties of a body affect the propagation in several ways. First, the high dielectric constant increases the electrical length of E-field antennas such as a dipole. Second, body tissue such as muscle is partly conductive and can absorb some of the signal. Additionally, it can also act as a parasitic radiator. This is significant when the physical antenna is much smaller than the optimum. Typical dielectric constant $\varepsilon_{r}$, conductivity $\rho$ and characteristic impedance $Z_{0}(\Omega)$ properties of muscle and fat are shown in Table \ref{tab:1}

\begin{table*}[!t]
\renewcommand{\arraystretch}{1.3}
\caption{Throughput and Power (in mW) of IEEE 802.15.4 and IEEE 802.11e under AC\textunderscore BE and AC\textunderscore VO \cite{36}}
\label{tab:2}
\centering
\begin{tabular}{|c|c|c|c|c|}
\cline{1-5}
& \multicolumn{1}{|c|}{} & \multicolumn{1}{|c|}{IEEE 802.15.4)} & \multicolumn{1}{|c|}{IEEE 802.11e (AC\textunderscore BE)
)} & \multicolumn{1}{|c|}{IEEE 802.11e (AC\textunderscore VO))} \\ \cline{1-5}

\multicolumn{1}{|c|}{Throughput} & Wave-form  & 100\% & 100\% & 100\%       \\ \cline{2-5}
\multicolumn{1}{|c|}{} & Parameter & 99.77\% & 100\% & 100\%       \\ \cline{1-5}
\multicolumn{1}{|c|}{Power (mW)} &  Wave-form & 1.82 & 4.01 & 3.57       \\ \cline{2-5}
\multicolumn{1}{|c|}{} &  Parameter & 0.26 & 2.88 & 2.77       \\ \cline{1-5}
\end{tabular}
\end{table*}

\subsection{Dipole Antenna}
For a dipole of length $10mm$, at 403MHz, the radiation resistance is $45m \Omega$ in air. The electrical length of the dipole is increased when surrounded by a material of high dielectric constant such as the body. 

\subsection{Loop Antenna}
For a loop of $10 mm$ diameter the area is $78.5 mm^{2}$. This gives the radiation resistance of $626 \mu \Omega$. However, the loop acts as a magnetic dipole, producing more magnetic field than a dipole. The loop is of use within the body as the magnetic field is less affected by the body tissue compared to a dipole or a patch and it can be readily integrated into existing structures. 

\subsection{Patch Antenna} 
\begin{figure}[!h]
\centering
\includegraphics[width=3in]{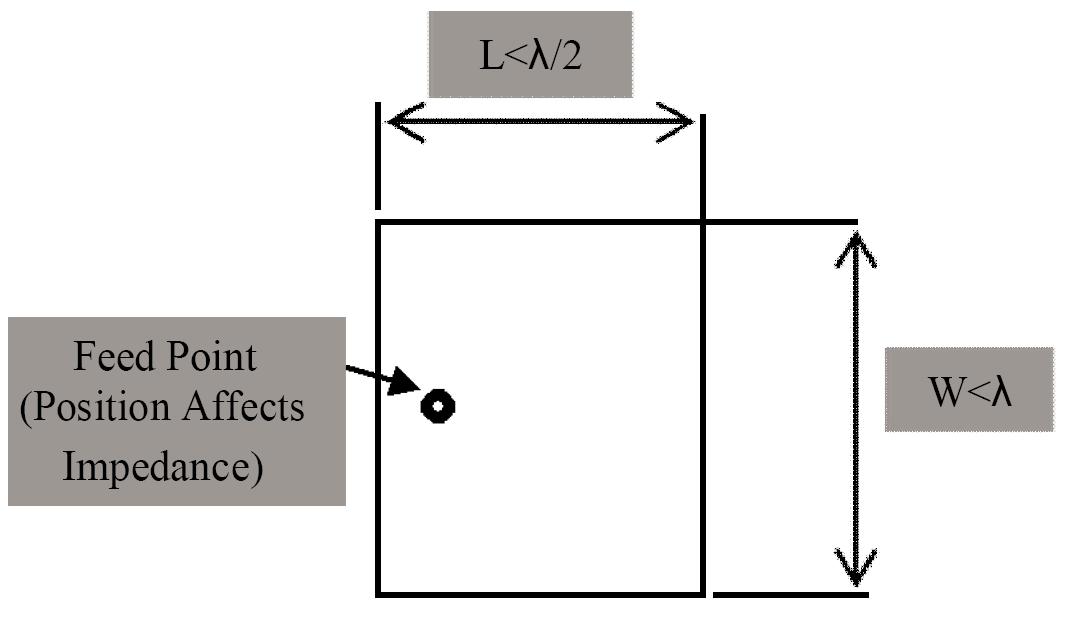}
\caption{Patch  antenna  plan  view, $\lambda$ in the surrounding medium} 
\label{fig:3}
\end{figure}

A patch antenna can be integrated into the surface of an implant. Without requiring much additional volume, the ideal patch will have dimensions as given in Fig. \ref{fig:3}. It acts as a $\lambda/2$ parallel plate transmission line with an impedance inversely proportional to the width. 

\begin{figure}[!h]
\centering
\includegraphics[width=3in]{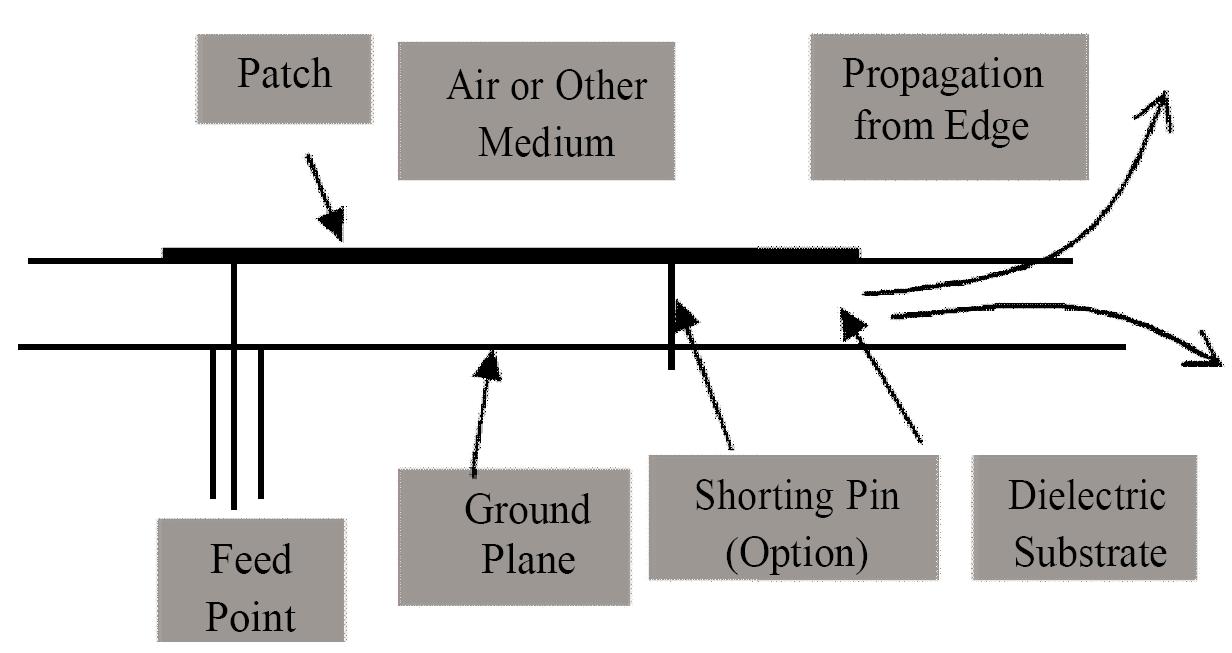}
\caption{Patch antenna side view} 
\label{fig:4}
\end{figure}

The  radiation  occurs  at  the  edges  of  the  patch as given in Fig. \ref{fig:4}. For in-body use, a full size patch is not an option. However, as it is immersed into the body tissue that has a dielectric constant in the order of 50, the electrical size of the patch becomes larger than would be in air. An electrically small patch will have low real impedance and therefore impaired performance compared to the ideal one. There are several other options such as Planar Inverted-F Antenna (PIFA), loaded PIFA, the bow tie, spiral and trailing wire. These antennas may have properties that may make them better suited for some applications. 

\subsection{Impedance Measurement}
The impedance of the patch and dipole is affected considerably by being surrounded by the body tissue. The doctor who fits it determines the position of an implant within a body. It may move within the body after fitting. Each body has a different shape with different proportions of fat and muscle that may change with time. This means that a definitive measurement of antenna impedance is of little value. Measuring it immersed in a body phantom can make an approximation of the impedance liquid \cite{20}. Using this impedance, the antenna-matching network can be designed with the provision of software controlled trimming as can be done with variable capacitors integrated into the transceiver. The trimming routine should be run on each power up or at regular intervals to maintain optimum performance.

\section{MAC Protocol}
The design and implementation of a low-power MAC protocol for WBAN is currently a hot research topic. The most challenging task is to accommodate the in-body nodes in a power-efficient manner. Unlike on-body nodes, the in-body nodes are implanted under human skin where the electrical properties of the body affect the signal propagations. Li et al. proposed a novel TDMA protocol for an on-body sensor network that exploits the biosignal features to perform TDMA synchronization and to improve the energy efficiency \cite{21}. Other protocols like WASP, CICADA, and BSN-MAC are proposed in \cite{22}-\cite{24}. The performance of a non-beacon IEEE 802.15.4 is investigated in \cite{25}, where the authors considered low upload/download rates mostly per hour. However, it has no reliable mechanism defined for on-demand and emergency traffic. 

\begin{figure}[!h]
\centering
\includegraphics[width=3in]{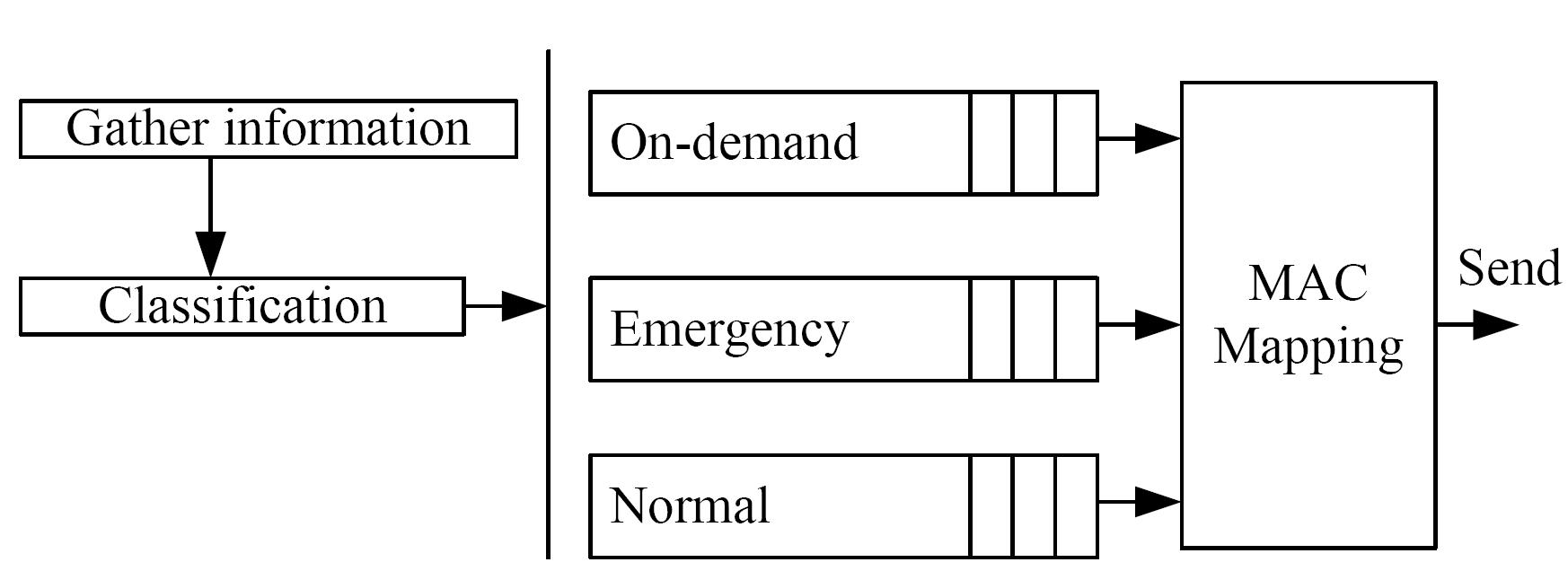}
\caption{WBAN MAC mapping} 
\label{fig:5}
\end{figure}

The WBAN traffic requires sophisticated low-power techniques to ensure safe and reliable operations. Existing MAC protocols such as SMAC \cite{26}, TMAC \cite{27}, IEEE 802.15.4 \cite{28}, and WiseMAC \cite{29} give limited answers to the heterogeneous WBAN traffic. The in-body nodes do not urge synchronized wakeup periods due to sporadic medical events. Medical data usually needs high priority and reliability than non-medical data. In case of emergency events, the nodes should access the channel in less than one second \cite{30}. IEEE 802.15.4 Guaranteed Time Slots (GTS) can be utilized to handle time critical events but they expire in case of low traffic load. Fig. \ref{fig:5} shows the required MAC mapping of the WBAN traffic.

The IEEE 802.15.4 can be considered for certain on-body sensor network applications but it does not achieve the required power level of in-body nodes. For critical and non-critical medical traffic, the IEEE 802.15.4 has several power consumption and QoS issues \cite{31}-\cite{34}. Also, it operates in 2.4 GHz band, which allows the possibilities of interference from other devices such as IEEE 802.11 and microwave. Dave et al. studied the energy efficiency and QoS performance of IEEE 802.15.4 and IEEE 802.11e \cite{35} MAC protocols under two generic applications: a wave-form real time stream and a real-time parameter measurement stream \cite{36}. Table \ref{tab:2} shows the throughput and the Power (in mW) for both applications. The AC\textunderscore BE and AC\textunderscore VO represent the access categories best-effort and voice in the IEEE 802.11e, respectively.

\begin{figure}[!h]
\centering
\includegraphics[width=3in]{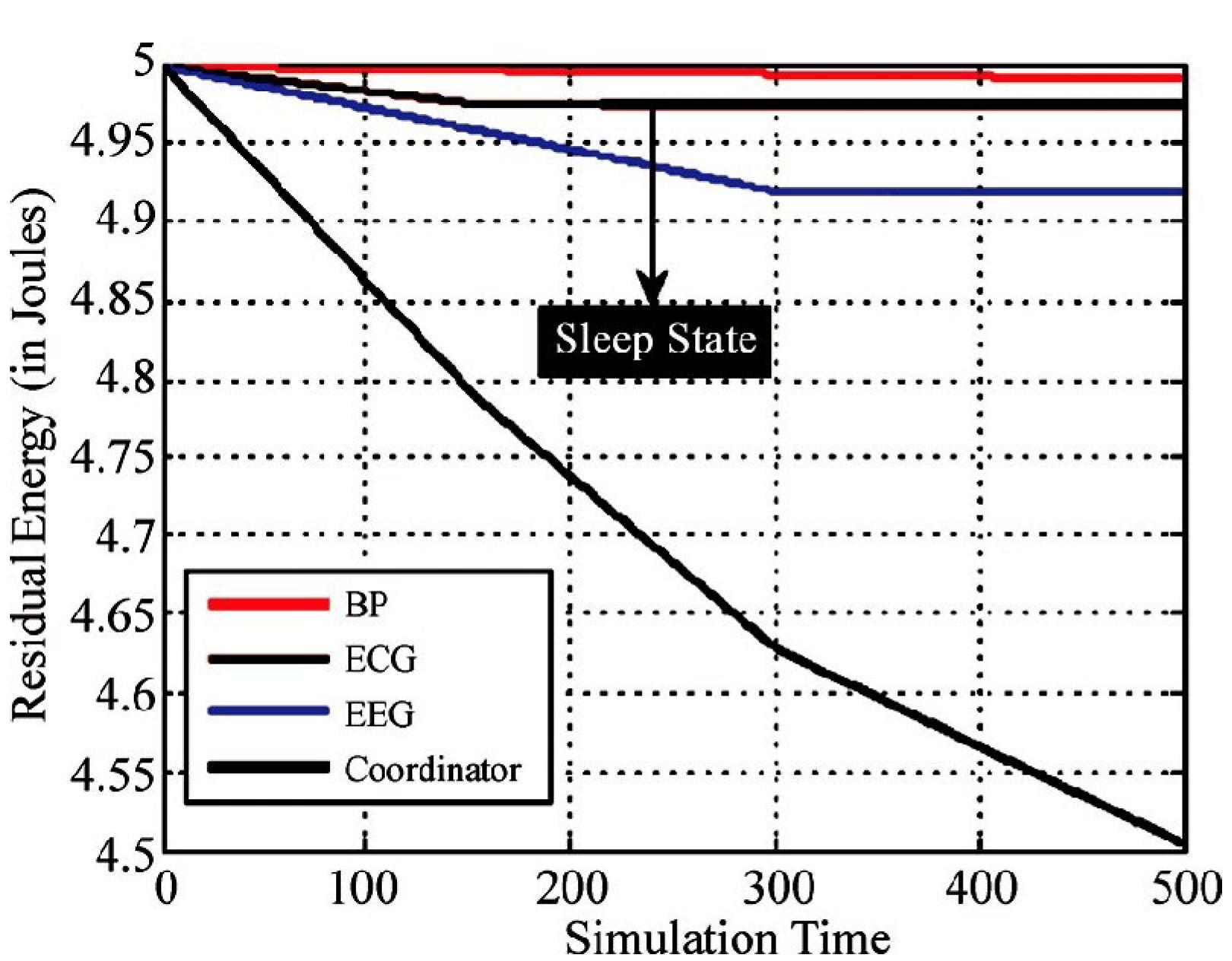}
\caption{Residual energy at on-body nodes} 
\label{fig:6}
\end{figure}

IEEE 802.15.4 uses CSMA/CA mechanism that does not provide reliable solutions for in-body nodes because the path loss inside human body results in improper Clear Channel Assessment (CCA). For a threshold of -85dBm and -95dBm, the on-body nodes cannot see the activity of in-body nodes when they are away at 3 meters distance from the body surface \cite{37}. An alternative solution is to use TDMA-based protocols for WBAN. We therefore analyzed the performance of a preamble-based TDMA \cite{38} protocol for an on-body sensor network. We used ns-2 \cite{39} for extensive simulations. Fig. \ref{fig:6} shows the residual energy at the on-body nodes and the coordinator. After the nodes finish their transmissions, they go into sleep mode. The Electrocardiogram (ECG) node sleeps after 150 seconds. When the Electroencephalography (EEG) node finishes its transmission at 300 seconds, the coordinator consumes less energy as indicated by the slight change in the curve. 

\begin{table*}[!t]
\renewcommand{\arraystretch}{1.3}
\caption{In-body and on-body sensor networks applications}
\label{tab:3}
\centering
\begin{tabular}{|p{2cm}|p{2cm}|p{1.7cm}|p{2cm}|p{2cm}|p{2cm}|p{1cm}|}
\cline{1-7}
Application Type & Sensor Node & Date Rate & Duty Cycle (per device)\% per time & Power Consumption & QoS (Sensitive to Latency) & Privacy \\ \cline{1-7}

\multicolumn{1}{|p{3cm}|}{In-body Applications} & Glucose Sensor & Few Kbps & $<$1\% & Extremely Low & Yes & High   \\ \cline{2-7}
\multicolumn{1}{|c|}{} & Pacemaker & Few Kbps & $<$1\% & Low & Yes & High   \\ \cline{2-7} 
\multicolumn{1}{|c|}{} & Endoscope Capsule & $>$ 2Mbps & $<$50\% & Low & Yes & Medium   \\ \cline{1-7} 
\multicolumn{1}{|c|}{On-body Medical Applications} & ECG & 3 Kbps & $<$10\% & Low & Yes & High   \\ \cline{2-7}
\multicolumn{1}{|c|}{} & SpO2 & 32 bps & $<$1\% & Low & Yes & High   \\ \cline{2-7}
\multicolumn{1}{|c|}{} & Blood Pressure & $<$10 bps & $<$1\% & High & Yes & High   \\ \cline{1-7}
\multicolumn{1}{|c|}{On-body Non-Medical Applications} & Music for Headsets & 1.4 Mbps & High & Relatively High & Yes & Low   \\ \cline{2-7}
\multicolumn{1}{|c|}{} & Forgotten Things Monitor & 256 Kbps & Medium & Low & No & Low   \\ \cline{2-7}
\multicolumn{1}{|c|}{} & Social Networking & $<$200 Kbps & $<$1\% & Low & No & High   \\ \cline{1-7}
\end{tabular}
\end{table*}

\begin{figure*}[!t]
\centering
\includegraphics[width=4in]{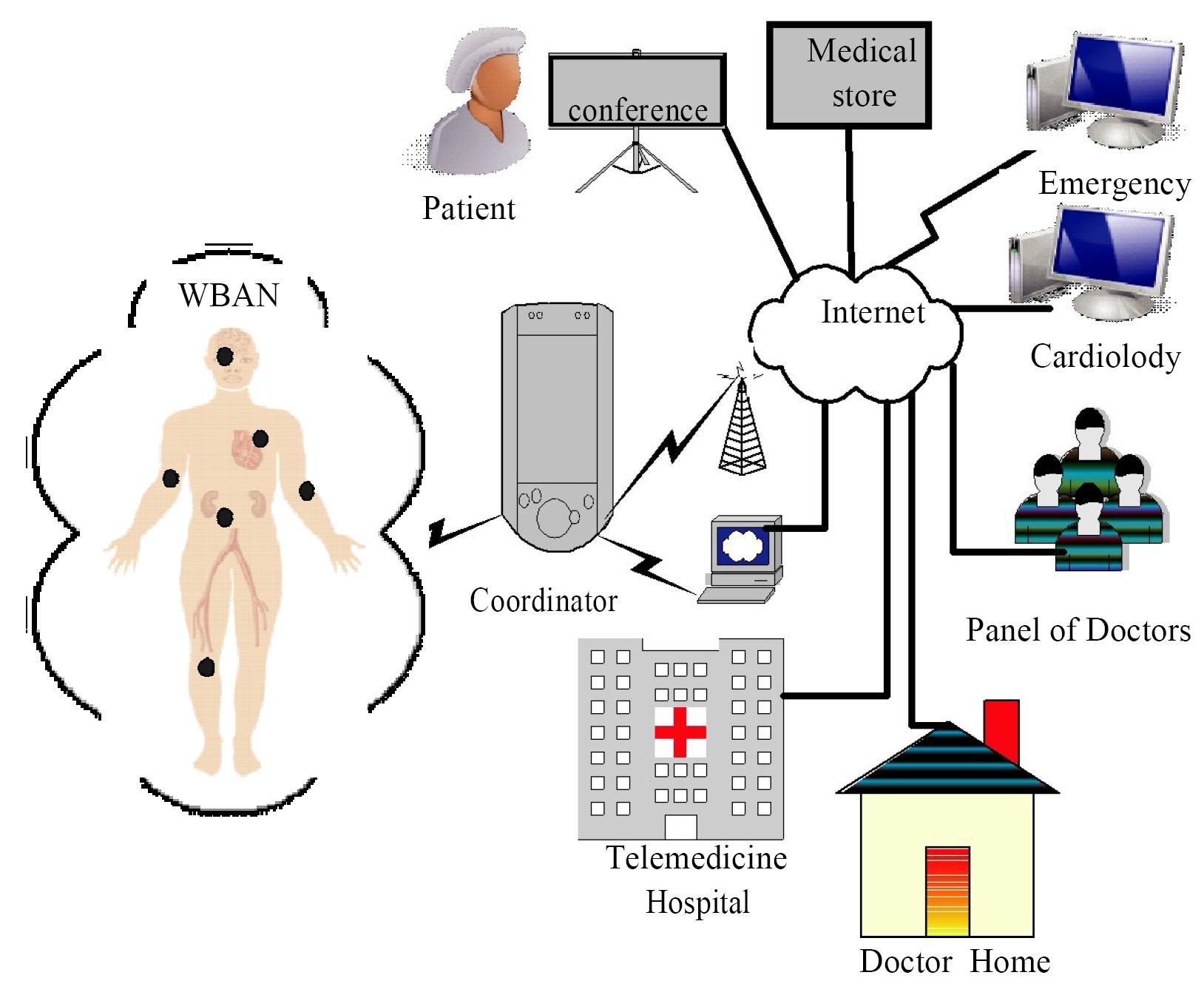}
\caption{A real-time telemedicine infrastructure for patient rehabilitation} 
\label{fig:7}

\end{figure*}\section{WBAN Applications}
WBANs have great potential for several applications including remote medical diagnosis, interactive gaming, and military applications. Table \ref{tab:3} shows some of the in-body and on-body applications \cite{40}. In-body applications include monitoring and program changes for pacemakers and implantable cardiac defibrillators, control of bladder function, and restoration of limb movement \cite{41}. On-body medical applications include monitoring heart rate, blood pressure, temperature, and respiration. On-body non-medical applications include monitoring forgotten things, establishing a social network, and assessing soldier fatigue and battle readiness. The following part discusses some of the WBAN applications:

\subsection{Cardiovascular Diseases}
A WBAN is a key technology to prevent the occurrence of myocardial infarction, monitors episodic events or any other abnormal condition and can be used for ambulatory health monitoring \cite{42}.

\subsection{Cancer Detection}
Cancer remains one of the biggest threats to the human life. According to National Center for Health Statistics, about 9 million people had cancer diagnosis in 1999 \cite{43} and this number increases every year. A set of miniaturised sensors capable of monitoring cancer cells can be seamlessly integrated in WBAN. This allows physician to diagnose tumors without biopsy. 

\subsection{Asthma}
A WBAN can help millions of patients suffering from asthma by monitoring allergic agents in the air and by providing real-time feedback to the physician. Chu et al proposed a GPS-based device that monitors environmental factors and triggers an alarm in case of detecting information allergic to the patient \cite{44}.

\subsection{Telemedicine Systems}
Existing telemedicine systems either use dedicated wireless channels to transfer information to the remote stations, or power demanding protocols such Bluetooth that are open to interference by other devices working in the same frequency band. These characteristics limit prolonged health monitoring. A WBAN can be integrated into a telemedicine system that supports unobtrusive ambulatory health monitoring for long period of time. Fig. \ref{fig:7} shows a real-time telemedicine infrastructure for patient rehabilitation.

\subsection{Artificial Retina}
Retina prosthesis chips can be implanted in the human eye, which assist patients (with limited or no vision) to see at an adequate level. 

\subsection{Battlefield}
WBANs can be used to connect soldiers in a battlefield and report their activities to the commander, i.e., running, firing, and digging. However, the soldiers should have a secure communication channel in order to prevent ambushes. 

\section{Conclusions}
In this paper, we reviewed the key aspects of WBAN including traffic classification, in-body antenna design, and MAC protocols. We provided a technical discussion on the in-body antenna design and supported patch antenna for in-body communication. We also discussed low-power MAC protocol for WBAN. We believe that existing low-power MAC protocols have several limitations to accommodate the heterogeneous traffic in a reliable manner and hence require new power-efficient solutions. We finally outlined some of the WBAN applications in ubiquitous healthcare sector.

\section*{Acknowledgment}
This research was supported by the The Ministry of Knowledge Economy (MKE), Korea, under the Information Technology Research Center (ITRC) support program supervised by the Institute for Information Technology Advancement (IITA) (IITA-2009-C1090-0902-0019).


\end{document}